\documentclass[reprint,amsmath,amssymb,aps,showkeys]{revtex4-1}

\pdfoutput=1
\usepackage{color}
\usepackage{verbatim}
\usepackage{amsmath}
\usepackage{amssymb}
\usepackage{bm}
\usepackage{latexsym}
\usepackage[normalem]{ulem}

\usepackage{epsfig} 
\usepackage{epstopdf}
\usepackage{graphicx}
\usepackage{hyperref}

\usepackage{lmodern}

\newcommand{\ba}{\begin{array}}
\newcommand{\ea}{\end{array}}
\newcommand{\be}{\begin{equation}}
\newcommand{\ee}{\end{equation}}
\newcommand{\bea}{\begin{eqnarray}}
\newcommand{\eea}{\end{eqnarray}}

\newcommand{\bfA}{{\mathbf{A}}}
\newcommand{\bfI}{{\mathbf{I}}}
\newcommand{\bfE}{{\mathbf{E}}}
\newcommand{\bftE}{{\mathbf{\tilde E}}}
\newcommand{\bfH}{{\mathbf{H}}}

\newcommand{\bfM}{{\mathbf{M}}}
\newcommand{\bfJ}{{\mathbf{J}}}

\newcommand{\bfk}{{\mathbf{k}}}
\newcommand{\bfr}{{\mathbf{r}}}
\newcommand{\bfF}{{\mathbf{F}}}
\newcommand{\bfG}{{\mathbf{G}}}
\newcommand{\bftG}{{\mathbf{\tilde G}}}

\newcommand{\bfd}{{\mathbf{d}}}
\newcommand{\bfe}{{\mathbf{e}}}
\newcommand{\bfp}{{\mathbf{p}}}
\newcommand{\bfx}{{\mathbf{x}}}

\DeclareMathAlphabet\mathbfcal{OMS}{cmsy}{b}{n}

\begin{document}


\title{Accurate 2.5D frequency domain radar waves modelling using weighted-averaging difference operators}

\author{Bernard Doyon}
 \email{bdoyon@cegepgarneau.ca}
\affiliation{
 Centre d'\'{E}tudes {N}ordiques (CEN),Universit\'e Laval, Qu\'ebec, Canada
}

\author{Bernard Giroux}
\affiliation{
Institut National de la Recherche Scientifique, Centre Eau Terre Environnement, Qu\'ebec, Canada
}


\begin{abstract}
Modelling radar wave propagation in frequency domain is appealing in full waveform inversion because it allows decreasing the non-linearity of the problem, decreasing the dimension of the data space, better description of attenuation, and handling efficiently multiple sources.  Besides, performing 2.5D modelling is interesting when physical properties can be assumed invariant in one horizontal dimension because it allows reducing drastically computation requirements  compared to the 3D case. In 2.5D, finite-difference methods can be used to propagate the wave in two directions in space and a spatial Fourier transform is performed in the third direction to get a full three dimensional solution.  With a simple central finite-difference implementation,  second order accuracy in space is obtained and up to twenty grid points per wavelength are necessary to accurately simulate electromagnetic waves. Such a large number of grid points will impact on the storage requirement associated with frequency domain modelling. We propose a high accuracy algorithm to solve the frequency domain electromagnetic wave equation by finite-differences in 2.5D. The algorithm relies on a nine-point stencil to build weighted-averaging numerical operators. The weights are chosen to minimize numerical dispersion and anisotropy, which allows relaxing the requirements on grid cell size and thus decreases computational costs by a factor of about $3.6$ compared to the central finite-difference method. This new algorithm reduces the numerical error without increasing the numerical bandwidth of the matrix system to solve, and can be easily transposed to 3D frequency domain modelling.

\end{abstract}

\keywords
{Algorithms; Computational methods;  Inverse problems; Numerical modelling; Ground penetrating radar} 

\maketitle


\section{Introduction}

	Modelling the propagation of electromagnetic waves is a very frequent step in the interpretation of ground penetrating radar (GPR) data. For instance, full waveform inversion techniques (FWI) can be performed  
if predictions of the radar traces at the receiving antenna can be made for a particular set of transmitters.  For most geologic media or in the case of FWI where the model is unknown, these predictions rely on solving Maxwell's equations on a spatial grid, either in the time domain or the frequency domain, since analytical methods are limited to homogeneous, layered, and waveguide models only. 

Modelling the propagation of waves  in the frequency domain (either viscoelastic waves for seismology or radar waves in the present context)  offers numerous advantages compared to time-domain approaches \cite{AJOFRANKLIN05,JO1996}. First, the stability problem is absent. Second, full waveform inversion procedures in the frequency domain can be performed only a few discrete frequencies \cite{PRATT1999}. Furthermore, wavefields from multiple sources can be rapidly obtained using direct solvers  because the factorization of the impedance matrix must be done only once \cite{PRATT1998}. Finally, the discrete frequency spectrum can be covered simultaneously with a parallel implementation.

A major drawback however is the amount of resources required to solve the underlying linear system, in particular when 3D problems are tackled. Indeed, once the frequency domain equations are discretized, the wave field at a given location for a particular frequency will be obtained by solving a sparse matrix system. Depending on the problem and the choice of solver (direct vs iterative), the memory complexity of the factorization of the impedance matrix and overall complexity may be very high, e.g. respectivelly $O(292N^4)$ and  $O(N^6)$ with $N$ the number of grid points in one dimension for the 3D viscoelastic case \cite{CLICHE2014,LI2015}. 
 
Common GPR profiling and crosshole experiments typically restrict sources and receivers to a plane and most of the time, variation of the medium properties normal to this plane can be neglected. For this reason, some authors have developed hybrid methods, where the frequency domain finite-difference is used to simulate wave propagation in two dimensions and where space Fourier transforms are performed in the
out-of-plane direction to complete the solution for the wave in the third spatial dimension. Such hybrid methods, with point source in a medium that varies in only two dimensions, are denoted {\em two and one half dimensional} (2.5D).  This modelling approach has been proposed for seismic waves \cite{SONG1995,ZHOU1998}. For airborne electromagnetics, Li et al. \cite{LI2016} developed a 2.5D frequency domain forward modelling algorithm, limited to low frequencies ($<$ 1 MHz, the usual range for airborne electromagnetic experiments) and where the displacement currents are neglected from the original Maxwell's equations. 

Recently, Ellefsen et al. \cite{ELLEFSEN2009} presented a 2.5D frequency domain forward modelling method for radar waves using central finite-differences on a staggered grid to discretize the wave equation, and pointed out the fact that the numerical dispersion associated with the central finite-differences would eventually limit the accuracy and applicability of the approach for high frequencies.

Increasing the order of the differential operator allows reducing the numerical dispersion but increases the numerical bandwidth of the sparse matrix, the key parameter affecting the memory required to solve the system \cite{STEKL1998}. For seismic modelling, optimal finite-difference operators were developed to increase accuracy and to limit the bandwidth. The idea behind these new operators is to average the finite-differences with the nearest neighbors on the grid, using weights chosen to minimize  numerical dispersion. This approach has been presented in 2D  for both the acoustic \cite{JO1996,SHIN1998} and viscoelastic equations \cite{STEKL1998,MIN2000}. To our knowledge, these optimal finite-difference operators have not been presented Maxwell's equations 2.5D.  

This paper fills the gap and describes the method that we developed to improve the accuracy and efficiency of 2.5D frequency domain radar wave modelling. Even if elastic waves equations share some similarities with the electromagnetic wave equations, the development of optimal finite-difference operators is not straightforward for the latter. Indeed, for radar wave in a conductive medium, the phase velocity depends on frequency and therefore, the medium presents some {\em intrinsic} dispersion that should not be associated with the numerical dispersion caused by finite differences. Approximations and choices must be made to build the optimal finite-difference operators in this context.

The paper is organized as follows. In the theory section, we first recall important aspects inherent to 2.5D frequency domain modelling of radar waves and present the new difference operators constructed from weighted average differences calculated with the nearest neighbors on the grid. 
 Dispersion analysis completes the theory section in order to obtain the weighting coefficients using a least-squares criterion.   We then analyze the numerical errors of the proposed
difference operators, first by comparing their associated dispersion curves to the standard second-order counterparts and second, by comparing the numerical solution with analytical results known for homogeneous and layered models.  
 Finally, we discuss briefly how to adapt the optimal difference operators for 3D radar wave modelling. It should be mentioned that we limited the content of this paper to study of finite-difference operators in the context of frequency domain modelling of radar waves. For phase and amplitude inversion of radar data using 2.5D frequency domain modelling, the reader is referred to \cite{ELLEFSEN2011b}.
 
 \section{Theory}
 
 {\bf \subsection{Outline of the 2.5D procedure}}

In this subsection, we mostly summarize the approach of Ellefsen et al. \cite{ELLEFSEN2009}, who gave a detailed description of 2.5D electromagnetic wave modelling. General considerations about the 2.5D approach may also be found in other references on seismic wave modelling \cite{PEDERSEN1994,SONG1995,FURUMURA1996,DOYON2014}.

The electric and magnetic field equations are solved in the frequency domain, for a rectangular Cartesian coordinate system whose horizontal and vertical directions are respectively given by the $x$ and $z$ axes.  Electromagnetic properties might depend upon frequency and may also vary in the $x$- and $z$- direction, but not in the $y$- direction. The transmitting antenna (the source) is an infinitesimal electric dipole and is modeled by a Dirac delta function. The dipole moment can have any orientation in space and can be located anywhere along $y$ within the $x-z$ limits of the modelling domain, but is usually placed at $y=0$. The electromagnetic fields solution obtained for this infinitesimal source is refer to as a {\em numerical Green's function}.

If we suppose a time dependent harmonic fields of the form $e^{-i\omega t}$, the 3D numerical Green's function is obtain by solving the following equation \cite{ELLEFSEN2009}
\be
\label{e_GeneralEeq}
\nabla \times \frac{1}{Z}(\nabla \times  \bfE) +Y \bfE = -\bfJ \,,
\ee
where $\bfJ$ is the electric current density (the infinitesimal electric dipole), $Z \equiv  -i \omega \mu$ and $Y \equiv -i \omega\epsilon_e$. With this notation, $\omega$ is the angular frequency, $\mu$ is the magnetic permeability and $\epsilon_e$ is the complex permittivity which depends on both the dielectric permittivity $\epsilon$ and the conductivity $\sigma$:
\[
\epsilon_e = \epsilon + \frac{i \sigma}{\omega}.
\]
The $Z$ and $Y$ terms will be respectively referred to as the {\em impedivity} and {\em admittivity} of the medium.

In equation~\eqref{e_GeneralEeq}, the electric field could be written as $\bfE = \bfE(x,z,y,\omega)$ to clearly show the three dimensional space dependency and to remind that the equation is solved in the frequency domain. To reduce this system of equation to a 2.5D configuration, we take the spatial Fourier transform of equation \eqref{e_GeneralEeq} with respect to the $y$ direction. The resulting equation for the transformed electric field (noted $\bftE=\bftE(x,z,k_y,\omega) $) now depends on the wavenumber $k_y$ and will be explicitly developed in the next section. It has to be solved for every $k_y$ of a predefined set of wavenumber values (to be specified), by discretizing  the model on the $x$-$z$ grid and by using finite-differences to replace the spatial derivatives. Using proper indexing, the discretization procedure will lead to a sparse matrix system of linear equations (see for instance \cite[Chapter 14]{INAN2011}). For an infinitesimal electric dipole of unit amplitude, this system can be written as

\be
\label{e_A}
\bfA \, \bftG =  -\frac{\boldsymbol{\Pi}(x,z)}{\Delta x_s \Delta z_s} \, ,
\ee
where $\bfA$ is a sparse, square and complex matrix,  $\bftG = \bftG(x,z,k_y, \omega)$ is the sought solution vector (the numerical Green's function of the model), $\boldsymbol{\Pi}(x,z)$ is the unit rectangular function approximating the infinitesimal electric unit dipole and ($\Delta x_s$, $\Delta z_s$) are the grid sizes at the source location.

The sparse matrix $\bfA$ contains the electric properties of the medium (impedivity $Z$ and admittivity $Y$). For a given frequency, this matrix has to be constructed for every wavenumber $k_y$, and each system of equation must be solved to get the numerical Green's function $\bftG$. The complete 3D solution for the Green's function is then obtained by the inverse Fourier transform:
\be
\label{e_GtildeInt}
\bfG(x,z,y,\omega) = \frac{1}{2\pi}\int_{-\infty}^{\infty}\bftG(x,z,k_y,\omega) e^{ik_yy}dk_y.
\ee

In practice, a small imaginary part is added to the angular frequency to stabilize the numerical integration of equation~\eqref{e_GtildeInt} \cite{SONG1995, BOUCHON2003,ZHOU2006,SINCLAIR2007}. The resulting complex frequency
\be
\label{e_omega_c}
\omega_c = \omega_r + i\omega_i,
\ee
is included in the impedance matrix and the solution $\bftG$ will then have an explicit dependency on this parameter. 

To avoid numerical reflection at the grid edges, the sparse matrix $\bfA$ is modified to include Perfectly  Matched Layers (PML) \cite{BERENGER1994,RAPPAPORT2000}. The PMLs must be adapted for the complex angular frequency \cite{ELLEFSEN2009}.  

The exact form of the sparse matrix $\bfA$ in equation \eqref{e_A} depends on the finite-difference scheme used to approximate the differential derivatives. 
In the next section, we present a new optimized finite-difference operator to solve the electric field equations in the 2.5D configuration. 

{\bf \subsection{A new finite-difference operator for 2.5D media}}

Using matrix notation, we write the homogeneous, source-free formulation of the Fourier transform of equation \eqref{e_GeneralEeq} as:
\be
\label{e_SourceFree}
\left [ \ba{ccc}
L_{11} & L_{12}  & L_{13} \\
L_{21} & L_{22}  & L_{23} \\
L_{31} & L_{32}  & L_{33}
\ea \right ]
\left [\ba{c} \tilde E^x \\ \tilde E^z \\ \tilde E^y \ea \right ] =\left [\ba{c} 0 \\ 0 \\ 0 \ea \right]
\ee
with
\[
L_{11}= Y - \frac 1 Z\frac{\partial^2}{\partial z^2}+ \frac{k_y^2}{Z}; \;\; \; \;   L_{12}=L_{21} = \frac 1 Z  \frac{\partial^2}{\partial x \partial z};
\]
\[
L_{22}=Y -\frac 1 Z \frac {\partial^2}{\partial x^2} + \frac{k_y^2}{Z}  ;  \;\; \; \; L_{13}=L_{31} = \frac{i k_y}{Z}\frac{\partial}{\partial x};
\]
\[
L_{33} =  Y- \frac 1 Z \frac {\partial^2}{\partial x^2} - \frac 1 Z \frac {\partial^2}{\partial z^2};\;\; \; \; L_{23}=L_{32} =\frac{i k_y}{Z} \frac{\partial}{\partial z}.
\]
We can derive the analytic phase velocity equation and the dispersion relation by assuming a harmonic expression for the electric field, of the form $\bftE = \bftE_0\exp(\bfk \cdot \bfr)$, with $\bftE_0$ the amplitude of the electric field at the origin and $\bfk = (k_x, k_z, k_y)$, the wavenumber. The non-zero solution for harmonic electric field requires the determinant of the matrix to be zero and leads to the dispersion relation for electromagnetic waves. With the definition of $Z$ and $Y$, this relation reduces to
\be
\label{e_kdispersion}
k^2 =  -YZ = \mu \epsilon \omega^2 + i \mu \sigma \omega.
\ee
For conductive media ($\sigma \ne 0$), the wavenumber is complex, resulting in some attenuation of the wave during propagation. Using $\beta$ for the real part of $k$ and $\alpha$ for the imaginary part, we can write the wavenumber as:
\be
\label{e_kgeneral}
k= \beta + i \alpha.
\ee
With the definition of the phase velocity, equations \eqref{e_kdispersion} and \eqref{e_kgeneral} can be combined to get a general expression for the analytic phase velocity
\be
\label{e_AnalyticPhaseVelo}
V \equiv \frac \omega \beta= \left[ \frac{\epsilon \mu}{2}\left(\sqrt{1+\frac{\sigma^2}{\epsilon^2 \omega^2}}+1\right)\right]^{-\frac{1}{2}}.
\ee
 The latter relation can be simplified depending on the ratio $\sigma/\omega \epsilon$.
 The so-called {\em lossless regime} corresponds to $\sigma = 0$. 

When using finite differences to compute the spatial derivatives in equation \eqref{e_SourceFree}, the numerical phase velocity (i.e. the resulting phase velocity on the discretized system) will depend on the finite difference operators. These operators should be chosen in such a way that the numerical phase velocity stays as close as possible to the analytical phase velocity. To find an expression for the phase velocity of the discretized system, we follow mostly \cite{CLICHE2014}, who gave this development for viscoelastic waves in a 3D media, but we modify and adapt the derivation for the 2.5D electromagnetic case presented in this paper.

The finite differences operators are obtained from values of the $\tilde E$ field at specific locations in the $x$-$z$ grid. An efficient discretization to formulate the finite-difference approximations is the {\em staggered grid}, in which the electric components $E_x, E_z$ and $E_y$ are estimated at different space locations. One such example is the well  known Yee grid, which was originally used for finite-difference time-domain simulations \cite{TAFLOVE2005} and which is also suitable in the frequency domain  \cite{CHAMPAGNE2001}.

In a medium discretized with a constant spatial step $\Delta= \Delta_x = \Delta_z$, equation \eqref{e_SourceFree} can be written as
\be
\label{e_SourceFreeDiscrete}
\left [ \ba{ccc}
a_{11} & a_{12}  & a_{13} \\
a_{21} & a_{22}  & a_{23} \\
a_{31} & a_{32}  & a_{33}
\ea \right ]
\left [\ba{c} \tilde E^x \\ \tilde E^z \\ \tilde E^y \ea \right ] =\left [\ba{c} 0 \\ 0 \\ 0 \ea \right]
\ee
with
\[
a_{11}= YD_m - \frac{1}{Z}\frac{D_{zz}}{\Delta^2}+ \frac{k_y^2}{Z};
\]
\[
a_{22}= YD_m - \frac{1}{Z}\frac{D_{xx}}{\Delta^2}+ \frac{k_y^2}{Z};
\]
\[
a_{33} =  YD_m - \frac{1}{Z}\frac{D_{xx}}{\Delta^2} - \frac{1}{Z}\frac{D_{zz}}{\Delta^2}
\]
\[
a_{12} =\frac{1}{Z}\frac {D_{xz}}{\Delta^2}; \; \; a_{21} = \frac{1}{Z}\frac{D^\star_{xz}}{\Delta^2}; \; \; a_{23}=\frac{i k_y}{Z}\frac{D_z}{\Delta}; 
\]
\[
a_{13} =\frac{i k_y}{Z}\frac{D_x}{\Delta}; \; \; a_{31} = \frac{i k_y}{Z}\frac{D^\star_x}{\Delta}; \; \; a_{32}= \frac{i k_y}{Z}\frac{D^\star_z}{\Delta}
\]
where $D_{xx}, D_{zz}, D_{xz}, D_x$ and $D_z$ are the finite-difference operators used to approximate the partial derivatives:
\[
\frac {\partial^2}{\partial z^2} \approx \frac{D_{zz}}{\Delta^2}; \;\;\;  \frac {\partial^2}{\partial x^2} \approx \frac{D_{xx}}{\Delta^2};\;\;\; \frac {\partial^2}{\partial x \partial z} \approx \frac{D_{xz}}{\Delta^2}
\]
\[
 \frac {\partial}{\partial x} \approx \frac{D_{x}}{\Delta}; \; \; \; \frac {\partial}{\partial z} \approx \frac{D_{z}}{\Delta}
\]
We have also introduced the three operators $D^\star_{xz}$, $D^\star_x$ and $D^\star_z$:
\[
 \frac {\partial^2}{\partial x \partial z} \approx \frac{D^\star_{xz}}{\Delta^2}
 \; \; \; \; \frac {\partial}{\partial x} \approx \frac{D^\star_{x}}{\Delta}; \; \; \; \frac {\partial}{\partial z} \approx \frac{D^\star_{z}}{\Delta}
\]
very similar to the non-star version ($D_{xz}, D_x$ and $D_z$). This distinction is made because we are using a staggered grid to build the operators. Since the operators act on different components of the field (for instance,  $D_x$ is applied on $\tilde E^y$ and $D^\star_x$ on $\tilde E^x$), this results in slightly different expressions for the operators. In equation \eqref{e_SourceFreeDiscrete}, we have also introduced the {\em lumped admittivity operator} $D_m$, linking the local effect of the $Y\bftE$ term to the neighboring points. For elastic wave propagation, this operator is known as the {\em lumped mass operator} and has proven useful to reduce the overall numerical dispersion \cite{JO1996}. We will give more details about this operator when specifying its explicit form in the next paragraphs.

\begin{figure}
\includegraphics[width=\columnwidth]{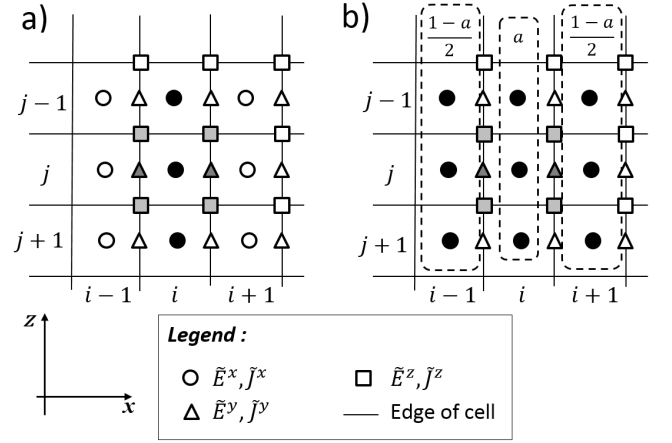}
\caption{\label{f_StaggeredGrid} Illustration of the stencil used to approximate partial derivatives with central finite-differences (Fig.~\ref{f_StaggeredGrid}a) and optimal finite-differences scheme (Fig.~\ref{f_StaggeredGrid}b).} 
\end{figure}

Fig.~\ref{f_StaggeredGrid} shows the staggered grid used to build the finite difference operators. Each cell is referenced with indices $i$ and $j$. The location of the components for the $\bftE$ field are identified with a different symbol ($\circ$ for $\tilde E^x$, $\triangle$ for $\tilde E^y$ and $\Box$  for $\tilde E^z$). Fig.~\ref{f_StaggeredGrid}a  illustrates the particular computational stencil implied with the first line of equation \eqref{e_SourceFreeDiscrete} for the standard central finite-difference scheme. The computational stencil is centered on the $\tilde J^x$ component and we grayed out the particular points needed to evaluate the finite-differences. For the other two lines of equation \eqref{e_SourceFreeDiscrete}, the stencil has to be moved and centered on $\tilde J^z$ (for the second line) and $\tilde J^y$ (for the third line).

\begin{table*}[t]
\caption{Central finite difference operators on a standard and extended staggered grid} 
\label{t_DiffOperators}
\centering 
\begin{tabular}{|c || c | c| } 
 \hline
 & Standard staggered grid & Extended staggered grid (9-point) \\ 
\hline \hline 
$[D_{zz}\tilde E^x]_{i,j}$ & $ \displaystyle \tilde E^x_{0,1}-2\tilde E^x_{0,0}+\tilde E^x_{0,-1}$ & $ a\left( \tilde E^x_{0,1}-2\tilde E^x_{0,0}+\tilde E^x_{0,-1} \right )  + \frac{(1-a)}{2} \sum_{\substack{s = -1\\s \ne 0}}^{s=1} \left( \tilde E^x_{s,1}-2\tilde E^x_{s,0}+\tilde E^x_{s,-1} \right )$    \\ 
$ [\displaystyle D_{xz}\tilde E^z]_{i,j}$ & $\displaystyle  \tilde E^z_{0,0}-\tilde E^z_{-1,0} -\tilde E^z_{0,-1}+\tilde E^z_{-1,-1}$ & (same as standard staggered grid)  \\[1ex]
$[\displaystyle  D_x\tilde E^y]_{i,j}$  & $\displaystyle  \tilde E^y_{0,0}-\tilde E^y_{-1,0} $ & (same as standard staggered grid)   \\[1ex]
$[D^\star_{xz}\tilde E^x]_{i,j}$& $\displaystyle   \tilde E^x_{1,1}-\tilde E^x_{1,0} -\tilde E^x_{0,1}+\tilde E^x_{0,0} $  & (same as standard staggered grid)    \\[1ex]
$[D_{xx}\tilde E^z]_{i,j}$ & $ \tilde E^z_{1,0}-2\tilde E^z_{0,0}+\tilde E^z_{-1,0}$  &  $ a \left( \tilde E^z_{1,0}-2\tilde E^z_{0,0}+\tilde E^z_{-1,0} \right ) + \frac{(1-a)}{2} \sum_{\substack{t = -1\\t \ne 0}}^{t=1}\left( \tilde E^z_{1,t}-2\tilde E^z_{0,t}+\tilde E^z_{-1,t} \right )$  \\
$[D_z \tilde E^y]_{i,j}$ & $\tilde E^y_{0,1}-\tilde E^y_{0,0}$ & (same as standard staggered grid) \\ [1ex]
$[D^\star_x\tilde E^x]_{i,j}$ & $\tilde E^x_{1,0}-\tilde E^x_{0,0}$ & (same as standard staggered grid) \\[1ex]
$[D^\star_z \tilde E^z]_{i,j}$  & $\tilde E^z_{0,0}-\tilde E^z_{0,-1}$ & (same as standard staggered grid) \\[1ex]
$[D_{xx} \tilde E^y]_{i,j}$ & $\tilde E^y_{1,0}-2\tilde E^y_{0,0}+\tilde E^y_{-1,0}$ & $a \left( \tilde E^y_{1,0}-2\tilde E^y_{0,0}+\tilde E^y_{-1,0} \right ) +\frac{(1-a)}{2}\sum_{\substack{t = -1\\t \ne 0}}^{t=1} \left( \tilde E^y_{1,t}-2\tilde E^y_{0,t}+\tilde E^y_{-1,t} \right )$\\
$[D_{zz} \tilde E^y]_{i,j}$& $ \tilde E^y_{0,1}-2\tilde E^y_{0,0}+\tilde E^y_{0,-1} $ & $a \left( \tilde E^y_{0,1}-2\tilde E^y_{0,0}+\tilde E^y_{0,-1} \right ) +\frac{(1-a)}{2} \sum_{\substack{s = -1\\s \ne 0}}^{s=1}\left( \tilde E^y_{s,1}-2\tilde E^y_{s,0}+\tilde E^y_{s,-1} \right )$  \\[1ex] 
\hline 
\end{tabular}
\label{table:nonlin} 
\end{table*}

\begin{table*}[t]
\caption{Lumped operator on a standard and extended staggered grid} 
\label{t_LumpOperator}
\centering 
\begin{tabular}{|c || c | c| } 
 \hline
 & Standard staggered grid & Extended staggered grid (5-point) \\ 
\hline \hline 
$[D_{m}\tilde E^\gamma]_{0,0}$ & $\displaystyle \tilde E^\gamma_{0,0}$ & $ b \tilde E^\gamma_{0,0} + \frac{(1-b)}{4} \left(\tilde E^\gamma_{0,1} + \tilde E^\gamma_{0,-1}+ \tilde E^\gamma_{1,0} + \tilde E^\gamma_{-1,0}\right)$    \\[1ex] 
\hline 
\end{tabular}
\label{table:nonlin} 
\end{table*}

The first column of Table~\ref{t_DiffOperators} gives the explicit form of the usual central finite-difference operators on the staggered grid for every partial derivative terms found in equation \eqref{e_SourceFreeDiscrete}. These operators suffer from numerical dispersion when the number of grid point per wavelength is too low \cite{ELLEFSEN2009}. In Table~\ref{t_DiffOperators}, we used the notation $\tilde E^\gamma_{0,0}$ with $\gamma = x, z,$ or $y$ components of $\bftE$ located on cell $i,j$ and  $\tilde E^\gamma_{\pm 1,\pm 1}$ for the same component on cell $i\pm1,j\pm1$, and half grid nodes indices are discarded to simplify the notation.

To reduce the numerical errors, a possible approach is to average the finite-differences with the nearest neighbors, using weights chosen to minimize numerical dispersion and ani\-so\-tro\-py. The simplest extension to the computational stencil is to consider a 9-point computational grid centered on the collocation point. For the approximation of spatial derivatives $\partial^2/\partial z^2$ and $\partial^2/\partial x^2$, our approach is to form $3$ finite-difference operators using 9 grid points and then average the operator using weighting coefficients.  Fig.~\ref{f_StaggeredGrid}b shows  the computational stencil implied in the first line of equation \eqref{e_SourceFreeDiscrete} with this 9-point difference operator. We first note that the mesh nodes used to approximate the operators for the $E^z$ and $E^y$ components are identical to the ones used for the standard central finite differential operator.  But for the second order partial derivative term $\partial^2E^x/\partial z^2$, the stencil is extended to a 9-point region (the black circle $\bullet$ in the figure). This term is now approximated with $3$ centered finite-differences. The first finite-difference is built with grid nodes taken from the same column as the collocation point. The other two finite-differences are built with grid nodes taken in the left-hand and right-hand columns. These three finite-difference operators are then averaged using weighting coefficient $a$ (for the first operator) and $(1-a)/2$ for the two others. The resulting explicit equation is

\begin{multline}
\left [ D_{zz}\tilde E^x \right]_{i,j} =  a\left( \tilde E^x_{0,1}-2\tilde E^x_{0,0}+\tilde E^x_{0,-1} \right )  \\ + \frac{(1-a)}{2} \sum_{\substack{s = -1\\s \ne 0}}^{s=1} \left( \tilde E^x_{s,1}-2\tilde E^x_{s,0}+\tilde E^x_{s,-1} \right ), \nonumber
\end{multline}

where the coefficient $a$ is chosen to minimize the numerical error. The second column of Table~\ref{t_DiffOperators} gives the explicit form of the difference operators for this simple extended 9-point stencil.

It is of course possible to propose other finite-difference operators, with more coefficients. We have to mention that the size of the computational stencil will affect the numerical bandwidth of the impedance matrix, which in turn will affect the computational time needed to solve the system.  Although it slightly increase the density of the impedance matrix (by a factor around $1.5$), the simple extension we propose does not change its bandwidth and reduces the numerical error by about $75\%$, as shown in the next section.  Thus, the computational time to compute the numerical Green's functions (up to a given error tolerance), is reduced with this new computational stencil since a larger grid spacing can be used.

We now turn our attention to the term $YD_m$ present in the diagonal elements of equation \eqref{e_SourceFreeDiscrete}. This term is normally approximated using local values (i.e.\ values at cell $i,j$) of $Y$,  $\tilde E^x, \tilde E^z$ and $\tilde E^y$. This is known as a consistent formulation, for which $D_m=1$. An alternative approach is obtained by interpolating the field values from the nearest nodes, where the interpolation is weighted by the local values of admitivity $Y$. This lump formulation is inspired from finite element method \cite{ZIENKIEWICZ2000} and has been used previously for seismic wave modelling \cite{MARFURT1984,JO1996}.  If we combine the consistent and lumped methods by a weighted average, the $D_m$ operator can be formulated (for homogeneous media) as:

\begin{multline}
\left[D_mE^\gamma \right]_{i,j} = b \tilde E^\gamma_{0,0} \\    \, \,  + \frac{(1-b)}{4} \left(\tilde E^\gamma_{0,1} + \tilde E^\gamma_{0,-1}+ \tilde E^\gamma_{1,0} + \tilde E^\gamma_{-1,0}\right), \nonumber
\end{multline}

with $\gamma = x, z,$ or $y$ and where the coefficient $b$ is also chosen to minimize the numerical errors. For operator $D_m$, we propose a simple five-point star, based on the work of \cite{STEKL1998} for viscoelastic modelling. This operator is also given in Table~\ref{t_LumpOperator} and compared to its standard formulation.

We now have to determine the optimal values of the two weight parameters $a$ and $b$. These two parameters are not independent and must be determined simultaneously. This is the topic of the next section.\\

{\bf \subsection{Optimization of weighting parameters}}

The new differencing and lump schemes, presented respectively in Tables \ref{t_DiffOperators} and  \ref{t_LumpOperator}, depend on parameters $a$ and $b$. These weighting coefficients control the amount of numerical error introduced by the finite-difference operators. To minimize this error, we look at the effect of the weighting coefficients on the numerical phase velocity and choose the set of coefficients for which the phase velocity is the closest to the known analytical expression.

The numerical phase velocity can be predicted in a standard fashion by assuming a plane-wave solution for the $\bftE$ field in the the discretized homogeneous source-free system given by equation \eqref{e_SourceFreeDiscrete}. With this assumption, the only non-zero solution for the field is found by making the matrix determinant of the resulting equation equal to zero. The general expression for this determinant leads to a polynomial of the sixth degree on the variable $\omega$, from which no analytical expression for the numerical phase velocity can be extracted easily. It is worth noting that the same procedure on the viscoelastic wave equations leads to a polynomial on the third degree on the variable $\omega^2$, polynomial from which the analytical expression for the numerical phase velocity can be extracted  and used to find the optimal coefficients \cite{CLICHE2014}. For radar waves, the conductivity $\sigma$ introduces a dispersion relation and the general constraint equation on the matrix determinant is more or less useful to get an expression for the numerical phase velocity. However, depending on dielectric regimes (the ratio $\sigma/\epsilon \omega$), the polynomial expression can be simplified and a corresponding numerical phase velocity can be deduced.
For the lossless regime, ($\sigma = 0$), the constraint on the determinant corresponds to a cubic polynomial on variable $\omega^2$ and leads to an analytical expression for the phase velocity on the discretized grid. We will use this particular regime to find the optimized coefficients and show in the next section that the coefficients are also efficient for regimes where $\sigma \ne 0$.

In appendix A, we present the mathematical details leading to the numerical phase velocity from the determinant of equation \eqref{e_SourceFreeDiscrete}. Two different expressions for the normalized phase velocity are obtained ($V^{\sigma = 0}_{P1}$ and $V^{\sigma = 0}_{P2}$), one for each of the transverse electromagnetic mode of propagation and are given by equation \eqref{e_V_Ph}. These expressions depend on $a$, $b$, $K$, $\theta$ and $\phi$, where $K$ is the wavenumber in grid point units ($K = k\Delta/2\pi$), $\theta$ is the propagation angle relative to the $z$ axis ($\theta = \{0 \ldots 2\pi \}$) and $\phi$ is the propagation angle relative to the $y$ direction ($\phi = \{0 \ldots \pi \}$):
\be
\label{e_VP1P2}
\ba{rcl}
V^{\sigma = 0}_{P1} (K,\theta, \phi, a, b), \\[2ex]
V^{\sigma = 0}_{P2} (K,\theta, \phi, a, b).
\ea
\ee
To determine the set of weighting coefficients that allows the normalized phase velocity to get as close as possible to unity, we use the iterative Levenberg-Marquardt method, which seeks to minimize in the least-squares sense, the difference between the expected value for the normalized velocity (i.e unity) and the values calculated by equation \eqref{e_V_Ph}. The reader is referred to \cite{CLICHE2014} for more details.

This procedure converges after a few iterations to a set of coefficients also given in Table~\ref{t_coeffs}. We will refer to this set as the optimal weighting coefficients.\\

\begin{table}[htb]
\centering 
\caption{\label{t_coeffs} Optimal weighted coefficients obtained by the optimization iterative algorithm} 
\resizebox{\columnwidth}{!}{
	\begin{tabular}{|c |c| } 
 	\hline
 	Initial value for optimization   & Optimal coefficients \\ 
	\hline \hline 
	a = 1.0 &  a = 0.9223    \\[1ex] 
	b = 1.0 & b = 0.7525 \\
	\hline 
	\end{tabular}
	}
\end{table}

\section{Error analysis}

{\bf \subsection{Numerical dispersion analysis}}

\begin{figure*}
\includegraphics[width=\textwidth]{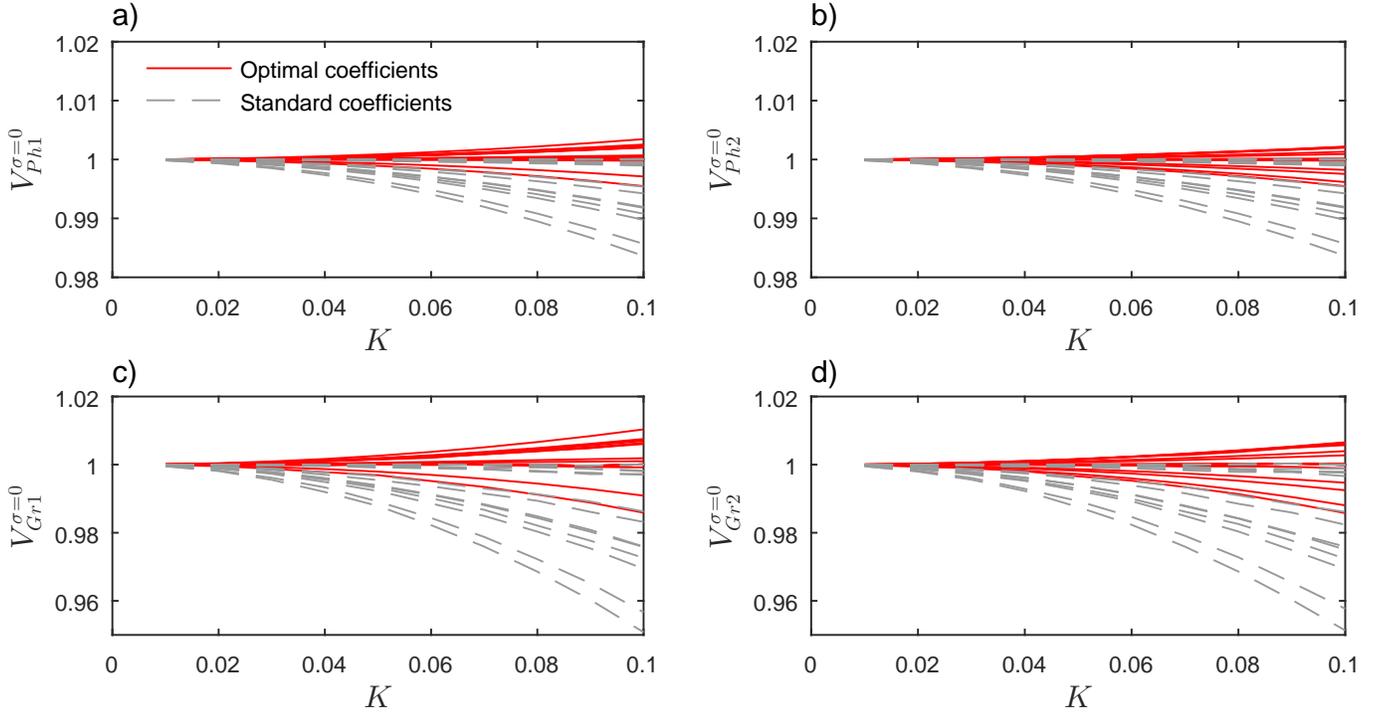}
\caption{\label{f_dispersion}Numerical phase velocities obtained with standard and optimal coefficients (indices $1$ and $2$ stands for two transverse modes) . Parameter $K$ represents the wavenumber in grid point unit. Several combinations of angles of propagation are plotted ($\phi = \{0, \pi/6, \pi/3,  \pi/2 \}$ and $\theta = \{0, \pi/12, \pi/6,  \pi/4 \}$).}
\end{figure*}

We can use the optimal weighting coefficients and examine the dispersion relation by plotting phase velocities for different propagation angles. Figs~\ref{f_dispersion}a and \ref{f_dispersion}b show the two normalized phase velocity (equation \eqref{e_V_Ph}) as a function of $K$.
For finite-difference time domain simulations with second-order accuracy in space, $20$ points per wavelength  are normally used for the shortest wavelength \cite{ELLEFSEN2009,TAFLOVE2005}, corresponding  to $K = 0.05$.  At this value of $K$, the error on the normalized phase velocity is around $1\%$ for certain angles of propagation when using standard coefficients (dashed lines on figure).  This error is reduced by about $75 \%$ with the optimal coefficients. It should be noted that even if the optimal coefficients were obtained by minimizing the error on phase velocity, these coefficients  will also reduce the error on group velocity (Figs~\ref{f_dispersion}c and \ref{f_dispersion}d).

{\bf \subsection{Accuracy of Green's functions}}

{\bf \subsubsection{Homogeneous model: low conductivity}}

We can use a homogeneous model to assess the accuracy of the numerical Green's functions calculated with the optimal weighting coefficients, by comparing the numerical and analytical solution. For this first numerical test, the infinitesimal electric dipole is oriented in the $z$-direction and located at the origin. The receiving antenna is located at $(x,y,z)=(-4,-0.1,0.1)$. The electromagnetic properties are isotropic and given in Table~\ref{t_Homo}, which also contains information about the parameters used to obtain the numerical solution for $46$ frequencies equally spaced between $0$ and $150$ MHz. It should be mentioned that the chosen parameters for this numerical model correspond to the simulation presented in \cite{ELLEFSEN2009}. We recall that the optimal weighting coefficient were obtained by assuming $\sigma$ = 0, and that for most of the frequency band used for this first numerical test, we are in a regime where $\sigma \ll \epsilon \omega$. (the low-loss dielectrics regime). We compare the electric field in the $z$-direction obtained numerically at the receiving antenna to the analytical expression. 

\begin{table}[htb]
\centering 
\caption{\label{t_Homo} Homogeneous model: Parameters for the numerical solution} 
\resizebox{\columnwidth}{!}{
	\begin{tabular}{|c |c| } 
 	\hline
 	Parameters   & Values \\
	\hline \hline 
	Dielectric permittivity ($\epsilon$) & $9 \, \epsilon_0$ \\
	Magnetic permeability ($\mu$) & $\mu_0$ \\
	Conductivity ($\sigma$) & 1.0 mS/m \\
	Source position & (x,y,z) = (0,0,0) m \\
	Receptor position & (x,y,z) = (4,-0.1,0.1) m \\
	Grid spacing $\Delta_x$ and $\Delta_z$ &  0.034 m   \\
	$\omega_r/2\pi$  (eq. \eqref{e_omega_c}) & 0 to 150 MHz (46 values) \\
	$\omega_i/2\pi$ (eq. \eqref{e_omega_c}) & 5 MHz \\
	Number of PML cells  & 10 at each boundary\\
	\hline 
	\end{tabular}
	}
\end{table}

The analytical solution for the isotropic non-conducting medium can be found in several reference textbooks on electromagnetic theory (see for instance \cite{STRATTON1941,CHENG1989}) 
and it is straightforward to transpose the solution to a more general dielectric media. For the electric dipole oscillating in the $z$ direction and placed at $z_{dipole}=h$ from the origin of an homogeneous dielectric media, the analytical solution for the $z$ component of the Green's function (noted $G^{Ho}_{zz}$) is:
\be
\label{e_GreenA}
G^{Ho}_{zz} = \frac{e^{ikr}}{4\pi (\sigma -i \epsilon \omega_c)r^3}\left(\frac{A_r(z-h)^2}{r^2} + B_r\right)
\ee
where
\[
A_r = 3 - 3ikr - k^2r^2       \, ; \quad B_r =-1+ikr + k^2r^2
\]
and with $r =\sqrt{x^2+y^2+(z-h)^2}$ and ($x$, $y$, $z$), the distances from the origin. 
The wavenumber $k$ has been given previously (equation \eqref{e_kdispersion}) and $\omega_c$ is the complex frequency used to obtain the numerical solution for the 2.5D geometry (equation \eqref{e_omega_c}).

The complete analytical Green's function given above is fairly complicated and it is common to examine it's behavior in regions near and far from the dipole \cite{STRATTON1941,CHENG1989}. In the region near the dipole (i.e.\ the {\em near zone}), $kr \ll 1$ and the leading term in equation \eqref{e_GreenA} is
\be
\bfG^{Ho}_{zz} \underset{kr \ll 1}\approx \frac{1}{4\pi (\sigma -i \epsilon \omega_c
)}
 		\left(\frac{3(z-h)^2-r^2}{r^5}\right),
\ee
which is the result one would obtain by an application of the laws of electrostatic. In this regime, the electric field is dominated by a reactive, non radiating part. In the last section, we assumed plane wave propagation to develop our optimal coefficients and we should not expect these coefficients to perform well in this near zone regime. On the other hand, in the {\em far zone}, (when $kr \gg 1$), the leading term in equation \eqref{e_GreenA} is
\be
\bfG^{Ho}_{zz} \underset{kr \gg 1}\approx \frac{k^2e^{ikr}}{4\pi (\sigma -i \epsilon \omega_c
)}
 		\left(\frac{x^2+y^2}{r^3} \right),
\ee
which has the same properties as those of a planar wave front. Therefore, the optimal weighting coefficients will be more appropriate for the latter regime.

\begin{figure*}[htb]
  \includegraphics[width=\textwidth]{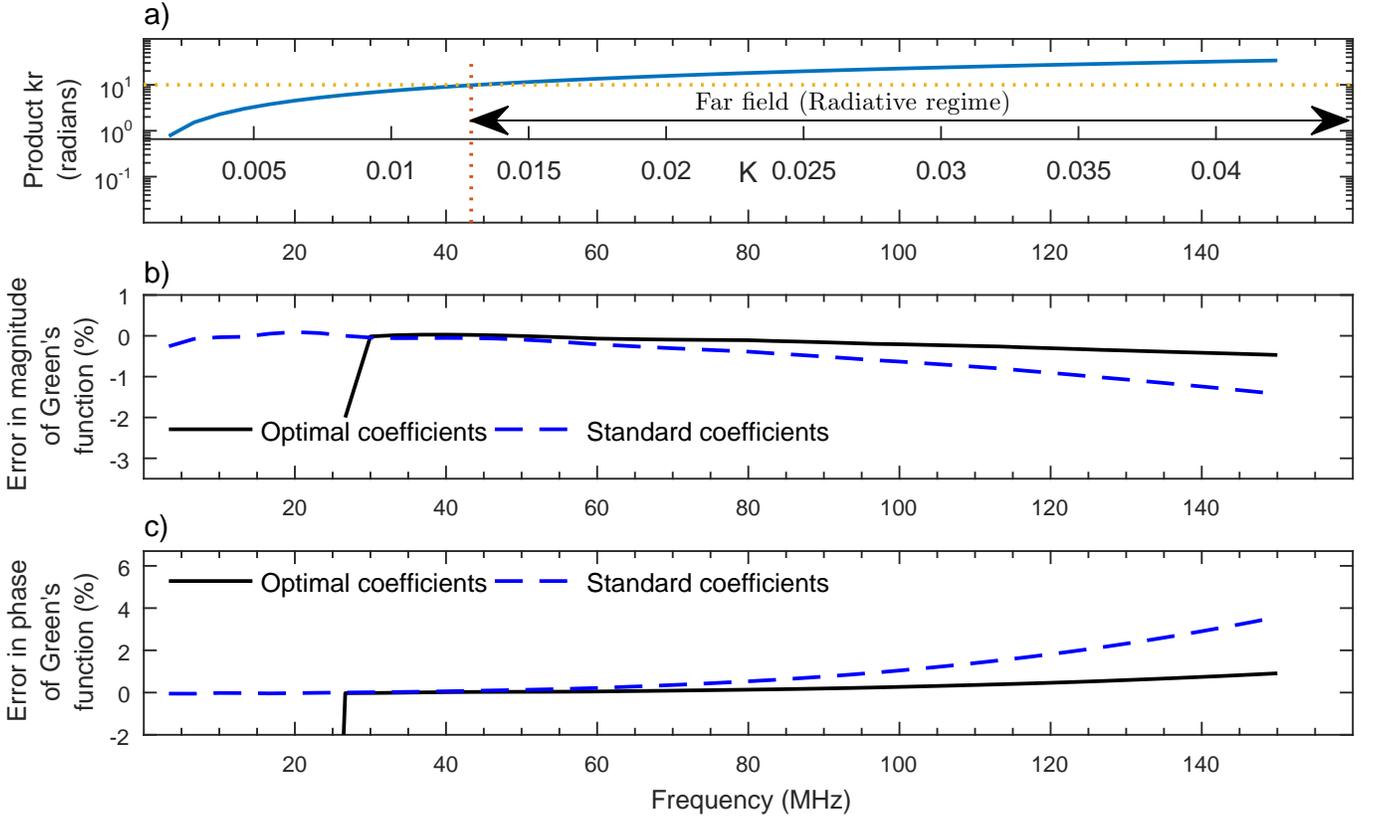}
  \caption{\label{f_ErrorLow} \small Accuracy of the Green's functions with standard and optimal coefficients for low conductivity media ($\sigma=10^{-3}$ S/m). Fig.~a shows the increasing value of the product $k \, r$ with the frequency (and the corresponding $K$ parameter). We consider the radiative regime for $k \,r>10$, a regime where the optimal coefficients are efficient. Figs~b and c illustrate the error for the magnitude and phase of the Green's functions calculated with standard and optimal coefficients. When the radiative condition is not met, the accuracy of the optimal coefficients is compromised (vertical drop around $f=30$ MHz).}
\end{figure*}

In Fig.~\ref{f_ErrorLow}a, we show the region where the radiative regime occurs assuming a cutoff at $kr = 10$.
Figs~\ref{f_ErrorLow}b and \ref{f_ErrorLow}c show respectively the error on the magnitude and phase of the Green's functions, quantified as
\be
\ba{rcl}
Error Magnitude& =& \displaystyle\frac{|G^{Ho}_{zz}| - |G_{zz}|}{|G^{Ho}_{zz}|}, \\[10pt]
Error Phase &=& \displaystyle \frac{\Phi_{G^{Ho}_{zz}} - \Phi_{G_{zz}}}{\pi }.
\ea
\ee

\begin{figure*}[htb]
  \includegraphics[width=\textwidth]{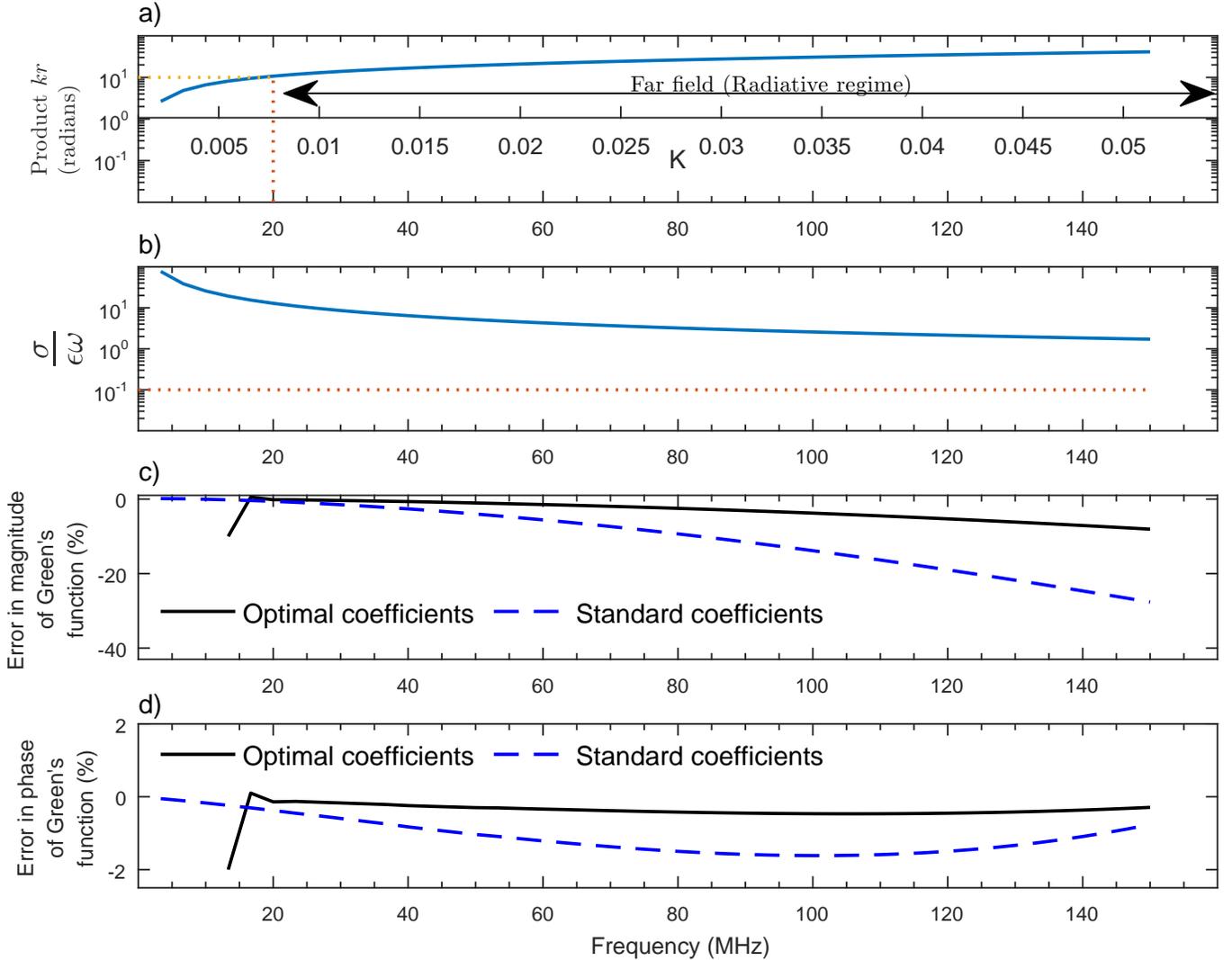}
  \caption{\label{f_ErrorHigh} \small Accuracy of the Green's functions with standard and optimal coefficients for high conductivity media ($\sigma=10^{-1}$ S/m). Figure b) shows the decreasing value $\sigma/\epsilon \omega$ with the frequency. For the selected value of $\sigma$, the low-loss dielectric condition is never met $\sigma/\epsilon \omega \ll 10^{-1}$. Figures c) and d) illustrate the error for the magnitude and phase of the Green's functions calculated with standard and optimal coefficients and Ellefsen's code. When the radiative condition is not met, the accuracy of the optimal coefficients is compromise (vertical drop around $f=15$ MHz).}
\end{figure*}

\begin{figure*}[htb]
\includegraphics[width=\textwidth]{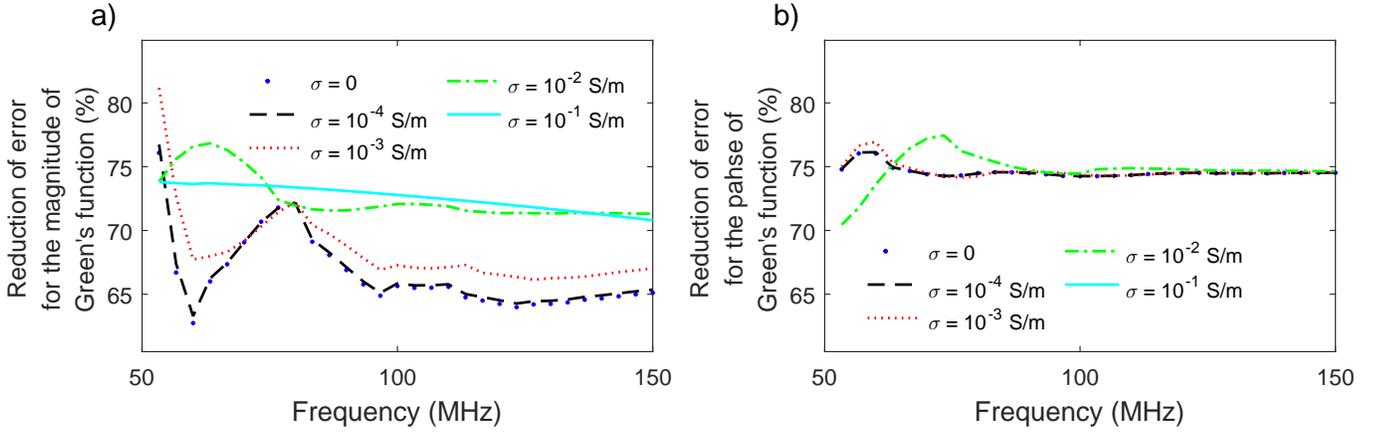}
\caption{\label{f_ReducError} \small Reduction of the error for the magnitude and phase of the Green's functions when using optimal coefficients compared to standard coefficients. The accuracy the optimal coefficients is not compromised in high conductivity media.}
\end{figure*}

\begin{figure*}[htb]
\includegraphics[width=\textwidth]{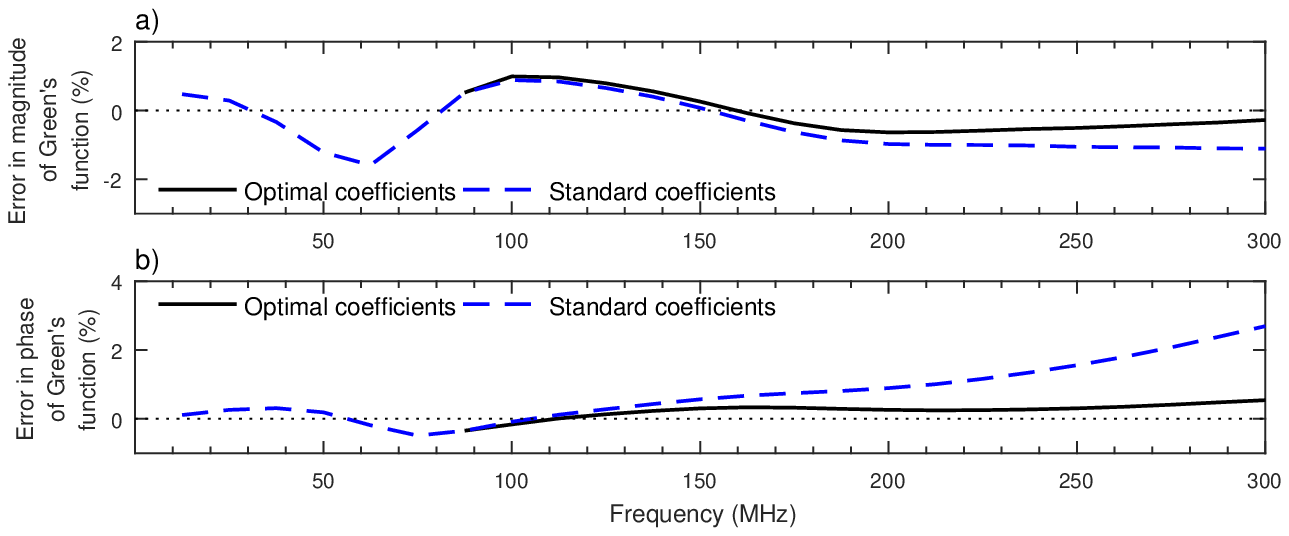}
\caption{\label{f_ErrorHete} \small Accuracy of the Green's functions for the layered model with standard and optimal coefficients. For reference, Green's functions were also calculated with Ellefsen's code.}
\end{figure*}

In the last two figures, the continuous and dashed lines represent the error when using respectively the optimal and the standard coefficients. We first note that the optimal coefficients reduce the error significantly as frequency increases. At 150 MHz, the wavenumber per grid point unit is approximately $K \approx 0.05$. As can be seen in the dispersion curves (Fig.~ \ref{f_dispersion}), for this value of $K$, the numerical dispersion introduces an error of about $0.5\%$ on phase velocity when using the standard coefficients, which translate in about $1.5\%$ error on the magnitude of the numerical Green's function. The optimal weighting coefficients becomes more useful as $K$ increases. On the other hand, for $K < 0.01$, the need for optimal coefficients is of less importance since numerical dispersion is negligible.
Furthermore, when $K$ is too small, the radiating regime condition is not met and the optimal coefficients becomes inefficient.

{\bf \subsubsection{Homogeneous model: high conductivity}}

Given that the optimal coefficient were developed assuming a zero conductivity in the medium, we have conducted a second experiment using a high conductivity ($\sigma$ = 0.1 S/m). All other parameters are the same as in Table~\ref{t_Homo}. For this conductivity and the tested frequency, the low-loss dielectric regime is never met (Fig.~\ref{f_ErrorHigh}b, the value $\sigma/\epsilon \omega$ is always greater than $10^{-1}$) and we could be tempted to think that the optimal coefficients would not be efficient. The errors on the Green's functions are presented in Figs~\ref{f_ErrorHigh}c and \ref{f_ErrorHigh}d. The optimal coefficients reduce the error on the Green's functions by more than $70 \%$ compared to the standard coefficients. For frequencies under 20 MHz $(K < 0.008)$, we note that the optimal coefficients become inefficient since the condition for a radiative regime is no longer valid.  

In order to quantify the improvement on the numerical Green's functions obtained with the optimal coefficients,  we evaluated the reduction of the error on the Green's functions (that is, the difference between the error obtained with the optimal coefficients and the standard coefficients, normalized by the error when using standard coefficients). We present in Fig.~\ref{f_ReducError} the reduction of the error on the Green's functions for $5$ different values of conductivity. All other parameters are given in Table~\ref{t_Homo}.  We plot the results for $f > 50$ MHz to be in the radiative regime for all conductivities.  The improvement obtained with the optimal coefficients for the magnitude of the Green's functions ranges between $65\%$ to $75 \%$ for the different tested conductivity (Fig.~\ref{f_ReducError}a). For the phase of the Green's functions, the improvement is around  $75\%$ and almost independent of the conductivity  (Fig.~ \ref{f_ReducError}b).\\

{\bf \subsubsection{Heterogeneous model: the layered model with high and low conductivities}}

The analytical development to obtain the optimal coefficient was performed with an homogeneous medium.  To evaluate the versatility of the approach, we conducted a numerical test with a heterogeneous model to see the performance of the coefficients if the conditions depart even further from the initial assumptions. We used  a 3-layer model to assess the accuracy of the numerical Green's functions. A saturated sand region of thickness $t=1$ meter in the $z$-direction is found between two clay layers (see \cite{ELLEFSEN2009}, fig. 2). The dipole is placed halfway in the sand region, is oriented in the $z$-direction, and the origin of the coordinate system is positioned at one of the clay-sand interfaces. The clay is found for $z>t$ and $z<0$ and the $z$ coordinate of the dipole is $z_{dipole} = t/2$. Table~\ref{t_Hete} contains the values of the parameters used for this simulation.

\begin{table}[htb]
\centering 
\caption{\label{t_Hete} Layered model: Parameters for the numerical solution} 
\resizebox{\columnwidth}{!}{
	\begin{tabular}{|c |c| } 
 	\hline
 	Parameters   & Values \\
	\hline \hline 
	Permittivity ($\epsilon$) & Sand=$40 \, \epsilon_0$; \,Clay=$20 \, \epsilon_0$  \\
	Permeability ($\mu$) &  Sand=$\mu_0$; \,Clay=$\mu_0$  \\
	Conductivity ($\sigma$) & Sand=$10^{-3}$ S/m;\,Clay=$0.5$ S/m  \\
	Source position & ($x$,$y$,$z$) = (0,0,$t$/2) m \\
	Receptor position & ($x$,$y$,$z$) = (1,-0.1,0.1) m \\
	Grid spacing $\Delta_x$ and $\Delta_z$ &  0.01 m   \\
	Thickness of sand layer & $t$ = 1 m (centered on source) \\
	Number of cells in z & 160 (Sand:100; Clay: 60) \\
	Number of cells in x  & 190 \\
	$\omega_R/2\pi$  (eq. \eqref{e_omega_c}) & 0 to 300 MHz (25 values) \\
	$\omega_I/2\pi$ (eq. \eqref{e_omega_c}) & 12.5 MHz \\
	Number of PML cells  & 10 at each boundary\\
	\hline 
	\end{tabular}
	}
\end{table}

For this model, the analytical solution for the electric field can be obtained from the Hertz potential \cite{TYRAS1969}. Three different Hertz potential functions containing arbitrary constants are imposed for each of the three regions and continuity conditions are used to obtain a unique solution. Details can be found in \cite{ELLEFSEN1999}\cite[p.~170]{TYRAS1969}.

Unfortunately,  both references contain typographic errors in the final stated solution and we felt the need to give the analytical solution in this paper. Since the receiver is placed in the sand region, we will give the analytical solution only for this central region. We use subscripts $s$ for sand and $c$ for clay. For a sand region of thickness $t$ and the electric dipole placed at a distance $h$ above one of the clay region interface (interface positioned at $z=0$)  and oscillating in the $z$ direction, the $z$ component of the analytical Green solution in the saturated sand region (noted $G^{Sand}_{zz}$) is
\be
\begin{split}
G^{Sand}_{zz} &= iM\int_0^\infty (A_\rho e^{ik_{s_z}z}+B_\rho e^{-ik_{s_z}z})J_0(k_{s_\rho}\rho) k_{s_\rho}^3dk_{s_\rho} \\
	& \quad \quad \quad \quad +G^{Ho}_{zz}
\end{split}
\ee
with
\[
\rho = \sqrt{x^2+y^2};  \, \, k_{s_z} = \sqrt{k_s^2 - k_{s_\rho}^2};  \, \, M =\frac{1}{4\pi (\sigma_s -i \epsilon_s \omega_c)}
\]
and
\[
A_\rho = R\left (\frac{1+Re^{i2k_{s_z}(t-h)}}{1-R^2 e^{i2k_{s_z} t}}\right ) \frac{e^{i k_{s_z} h}}{k_{s_z}},
\]
\[
B_\rho = Re^{2k_{s_z} b}\left (\frac{1+Re^{i2k_{s_z}(h)}}{1-R^2e^{i2k_{s_z} t}}\right ) \frac{e^{-i k_{s_z} h}}{k_{s_z}},
\]
where
\[
R = \frac{(k_c/k_s)^2k_{s_z} - k_{c_z}}{(k_c/k_s)^2k_{s_z} + k_{c_z}}.
\]
In this last equation, $G^{Ho}_{zz}$ is the solution for a homogeneous sand medium (equation \eqref{e_GreenA}), $J_0$ is the Bessel functions of the first kind and the integral represents the correction due to multiple reflection coming from the clay regions. Even if this integral has multiple poles, it can be evaluated by direct summation because the complex frequencies (equation \ref{e_omega_c}) introduce complex values for the wavenumber. The imaginary parts of the wavenumber add some numerical damping to the summation \cite{BOUCHON2003} which prevents the divergence of the summation around the poles.

Fig.~\ref{f_ErrorHete} shows the error on the magnitude and phase of the numerical Green's functions calculated both with the optimal weighting coefficients and the standard coefficients. The accuracy of the numerical Green's functions is improved with the use of the optimal coefficients, especially as frequency increases. Therefore, the optimal coefficients reduce the numerical dispersion even for a heterogeneous medium. In practice, this implies that a coarser grid spacing can be used for numerical simulation without conceding on accuracy.\\

{\bf \subsection{Computational time considerations}}

In the preceding section, we showed that the numerical error on the calculated Green's functions could be reduced by using a weight averaging finite-difference method. To evaluate the benefits of using the optimized coefficients algorithm, we used a computational time criteria and proceeded the following way:  First, we used standard coefficients to calculate, for the homogeneous media given in table \ref{t_Homo}, the numerical Green's functions for different frequencies and for a particular grid spacing. We next evaluated the error for each frequency with the help of the analytical Green's functions. We then used the optimal coefficients to calculate another set of numerical Green's functions, adjusting the grid spacing by trial and error to match the error obtained with the standard coefficients. We conducted this experience for different conductivity values. In general, for a similar numerical error, the grid spacing for the optimized coefficients was about twice as large than the grid spacing for the standard coefficients. We finally measured the time to calculate the numerical Green's functions for different scenarios based on the grid sizes (i.e. number of grid points) for standard and optimized coefficients. For a given scenario and in order to get the same numerical error for both finite differences methods, the grid spacing was doubled for the optimized coefficients (i.e. $\Delta_{Opt} = 2\Delta_{Std}$).   Therefore, for every tested scenarios, the total number of grid points was always reduced by a factor of $4$ with the optimal coefficients compared to the standard coefficients.  To minimize the fluctuations in the timing results, $20$ simulations were run for every grid size scenarios.

The average of these $20$ simulations for a particular scenario (corresponding to $50 \, 861$ grid points for the standard coefficients), are presented in table \ref{t_CompuTime}. The total time to compute the Green's function in the space domain for each frequency is given. To obtain the Green's function in the space domain for a given frequency, we recall that several Green's functions in the $k_y$ wavenumber domain need to be calculated. Therefore, the total time presented in table \ref{t_CompuTime} represents the time to fill the impedance matrices for the different wavenumber values, the time to solve the sparse matrix systems in the wavenumber domain (i.e. equation \ref{e_A}, which is the limiting operation), and the time to get the solution in the space domain (i.e. solving equation \eqref{e_GtildeInt} by direct summation). All the codes were written within the MATLAB environment  \cite{MATLAB2017}. The solution for the sparse system of equations is obtain with a direct solver (i.e. the UMFPACK library \cite{DAVIS2011} included with MATLAB). It should be mentioned that direct solvers are well adapted for multiple sources configuration often encountered in FWI problems since solutions for multiple sources can be rapidly obtained once the impedance matrix has been factorized. Computations were done on a dedicated server running CentOS linux 7 with 256 GB of RAM and four 3.3 GHz 8-core Intel Xeon E5-4627 v2 processor having each 16 MB of cache.

From table \ref{t_CompuTime}, we note that the average time increases with the frequency, since the integration interval of equation \ref{e_GtildeInt} (which is solved by direct summation), depends on the frequency \cite{ELLEFSEN2009,BOUCHON2003}. For every frequency, the finite difference method using optimal coefficients is about $3.6$ times faster than that using standard coefficients. The average time and average ratio per frequency  is given in the last line of the table \ref{t_CompuTime}.

\begin{table}
\centering 
\caption{\label{t_CompuTime} Computational time to compute Green's functions using standard and optimal coefficients (number of grid points for the standard coefficients: $50861$) }
\resizebox{\columnwidth}{!}{
	\begin{tabular}{|c |c|c|c| } 
 	\hline
 	Frequency & Standard coefficients & Optimal coefficients & Time Ratio  \\
 	 		& with $\Delta_{Std} =0.03$ & $\Delta_{Opt} =2\Delta_{Std} $ &   \\
      (MHz) & (Seconds) & (Seconds) & \\
	\hline \hline 
30 & 27.2 & 7.8  & 3.5 \\
40 & 31.1 & 8.8 & 3.5 \\
50 & 35.1 & 10.0  & 3.5 \\
60 & 39.1 & 11.1 & 3.5 \\
70 & 43.2 & 13.4 & 3.2 \\
80 & 51.1 & 14.4  & 3.5 \\
90 & 55.1& 15.6 & 3.5 \\
100 & 59.2 & 16.7  & 3.5 \\
110 & 63.2   & 17.9  & 3.5 \\
120& 67.2  & 20.1  & 3.3 \\
130& 75.2  & 21.3  & 3.5\\
140& 79.3 & 22.3  & 3.5\\
150& 83.1 & 23.5  & 3.5\\
	\hline 
Average: &54.5 &15.6 & 3.5 \\
\hline
	\end{tabular}
	}
\end{table}

\begin{figure}
\includegraphics[width=\columnwidth]{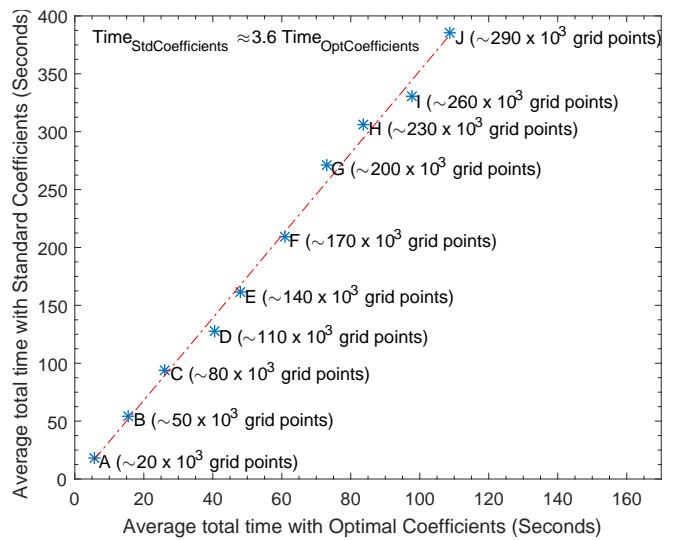}
\caption{\label{f_Chrono} Average time per frequency to compute the numerical Green's functions for different number of grid points scenarios. With the optimal coefficients algorithm, the time needed for calculation is reduced by a factor of about $3.6$ compared to the standard coefficients finite difference method. Each scenario is labeled with letter A to J, corresponding to an increasing number of grid points. The number of grid points for each scenario is given on the graph for the standard coefficients algorithm. The number of grid points for the optimal coefficients algorithm is reduced by a factor of $4$ (corresponding to a grid spacing larger by a factor of $2$) to obtain the same numerical error on the calculated Green's functions than that obtained with the standard coefficients.}
\end{figure}

Figure \ref{f_Chrono} shows the average time per frequency for the different scenarios performed. The scenarios are  labeled with letter A to J, corresponding to an increasing number of grid points (The number of grid points for each simulations is given in the graph for the standard coefficients). As we mentioned above, the number of grid points for the optimal coefficients algorithm is reduced by a factor of $4$ (corresponding to a grid spacing larger by a factor of $2$) to obtain the same numerical error on the calculated Green's functions than that obtained with the standard coefficients. For the tested grid sizes, the finite difference method using optimal coefficients is in general about $3.6$ times faster than that using standard coefficients. \\											

\section{Discussion and conclusion}

We have presented in this paper a method to obtain optimal weighting coefficients to compute finite-difference operators in order to solve the frequency-domain electromagnetic wave equations in a heterogeneous 2.5D domain. Even if the optimization approach used to obtain the coefficients was performed with lossless media ($\sigma = 0$), we showed that this assumption does not affect their performance and that they can be efficient in conductive media. Moreover, using the optimal coefficients allows reducing the computation time by a factor of at least 3.

The optimal weighting coefficients can be seen as a simple extension of the standard stencil used for finite differentiation. The difference operators we proposed are constructed with a 9-point stencil. Future work should be done to explore other expressions for these difference operators.  For instance, a larger stencil could be used to write the operators $D_{xx}$, $D_{zz}$ and $D_{yy}$ (using $25$ points). On this larger stencil, other expressions for $D_{xz}$, $D_x$ and $D_z$ could also be proposed. Recall that for a given grid spacing, a larger stencil would probably increase the accuracy of the solution but would also increase the computational time. A quantitative analysis is needed to assess if accuracy improvement using a larger stencil compensate the cost in computational time. 

It should also be mentioned that the optimal weighting coefficients proposed in this contribution were developed for a $2.5$D domain. It is straightforward to modify the method in order to get the optimal coefficients for a $3$D domain. To do so, some coefficients in equation \eqref{e_SourceFreeDiscrete} would have to be modified:
\[
a_{11} \rightarrow YD_m - \frac{1}{Z}\frac{D_{zz}}{\Delta^2}- \frac{D_{yy}}{Z\Delta^2};
\]
\[
a_{22}\rightarrow YD_m - \frac{1}{Z}\frac{D_{xx}}{\Delta^2}- \frac{D_{yy}}{Z\Delta^2}
\]
\[
a_{13} \rightarrow \frac {D_{xy}}{Z\Delta^2}; \; a_{31} \rightarrow \frac{D^\star_{xy}}{Z\Delta^2}; \; a_{23} \rightarrow \frac {D_{zy}}{Z\Delta^2}; \;  a_{32}\rightarrow \frac{D^\star_{zy}}{Z\Delta^2}
\]
and the extended and lumped operators given in Table~\ref{t_DiffOperators} and Table~\ref{t_LumpOperator} would have to be generalized for a $3$ dimensional stencil (see \cite{CLICHE2014}).

\section*{Computer code availability}
A MATLAB program to solve equations \ref{e_A} and \ref{e_GtildeInt} has been prepared and can be found on GitHub: \url{https://github.com/BernardDoyon/2_5DGreenFunctionRadar}. The {\em main.m} file can be use to specify the modelling parameters (e.g. size of domain, position of sources and receptors, frequencies, etc). More information about the program can be found in the README.md file.

\section*{Acknowledgements}
We are grateful the financial support from the Fonds de Re\-cher\-che
Nature et Technologies (FRQNT, Qu\'ebec) to B.D. and Natural Sciences and Engineering
Research Council of Canada (NSERC) to B.G (Discovery Grant RGPIN-2017-06215 ).

\bibliographystyle{seg_doi} 
\bibliography{DoyonGiroux2020} 

\newpage

\appendix

\section{Normalized phase velocities for finite-difference operator: the lossless regime case}

 We present in this appendix the mathematical details leading to the phase velocity, from the matrix determinant of equation \eqref{e_SourceFreeDiscrete}. For the lossless regime ($\sigma = 0)$, this equation is:
\be
\label{e_SourceFreeSigma0}
\left [ \ba{ccc}
a_{11}^{\sigma=0}& a_{12}  & a_{13} \\
a_{21} & a_{22}^{\sigma=0}  & a_{23} \\
a_{31} & a_{32}  & a_{33}^{\sigma=0}
\ea \right ]_{\sigma=0}
\left [\ba{c} \tilde E^x \\ \tilde E^z \\ \tilde E^y \ea \right ] =\left [\ba{c} 0 \\ 0 \\ 0 \ea \right]
\ee
with
\[
a_{11}^{\sigma=0}= -\omega^2 \mu \epsilon D_m - \frac{D_{zz}}{\Delta^2}+ k_y^2;
\]
\[
a_{22}^{\sigma=0}= -\omega^2 \mu \epsilon D_m - \frac{D_{xx}}{\Delta^2}+ k_y^2
\]
\[
a_{33}^{\sigma=0} =  -\omega^2 \mu \epsilon D_m - \frac{D_{xx}}{\Delta^2} - \frac{D_{zz}}{\Delta^2}
\]
and all other elements $a_{ij}$ are given below equation \eqref{e_SourceFreeDiscrete}.
By substituting an harmonic vector solution on the discretized grid, e.g.
\be
\label{e_HarmonicFunct}
\bftE_{i+s,j+t} = \bftE_{i,j}e^{i(k_xs\Delta + k_zt \Delta)}
\ee
where $(k_x, k_z)$ is the wave vector in equation \eqref{e_SourceFreeSigma0}, one obtains a linear homogeneous system of 3 equations and 3 unknowns ($\tilde E^x_{i,j}, \tilde E^z_{i,j}$ and $\tilde E^y_{i,j}$). For a non zero solution for the unknowns, the determinant of the matrix must be zero, leading to a cubic polynomial on variable $\omega^2$ that we write in the following way:
\be
\frac{A}{[D_m\epsilon \mu \Delta^2]^3} +\frac{B[\omega^2]}{[D_m\epsilon \mu \Delta^2]^2} + \frac{C [\omega^2]^2 }{[D_m\epsilon \mu \Delta^2]}+ [\omega^2]^3 = 0
\ee
with $A, B$ and $C$ given by:
\be
\label{e_ABC}
\ba{rcl}
A &=&  k_y^4\Delta^4\left (D_{xx} + D_{zz} - D_xD_x^\star  - D_zD_z^\star\right )  \\
	& & + k_y^2\Delta^2 \left(D_xD_x^\star D_{xx}  + D_zD_z^\star D_{zz} - D_{xx}^2-D_{zz}^2 \right . \\
	& & \hspace{0.5in} \left . + D_x^\star D_zD_{xz} + D_xD_z^\star D_{xz}^\star -2 D_{xx} D_{zz}\right )\\
	& &\hspace{0.1in} + D_{xx}^2 D_{zz} + D_{xx}D_{zz}^2 - D_{xz}D_{xz}^\star(D_{xx} + D_{zz})   \\[1ex]
B &=& k_y^4 \Delta^4 + k_y^2\Delta^2(D_xD^\star_x + D_zD^\star_z - 3D_{xx}-3D_{zz})\\
& & \; \; \; \;+ D_{xx}^2 + D_{zz}^2  + 3 D_{xx}D_{zz} - D_{xz}D^\star_{xz}\\ [1ex]
C&=&2(D_{xx} + D_{zz} -  k_y^2\Delta^2).
\ea
\ee

The roots of this cubic polynomial are
\be
\ba{rcl}
\omega_1 &=& \displaystyle \frac{1}{\Delta \sqrt{\epsilon \mu}}\sqrt{\frac{-\frac{C}{3}+ (S+T)}{D_m}} \\
\omega_2 & = & \displaystyle  \frac{1}{\Delta \sqrt{\epsilon \mu}}\sqrt{\frac{-\frac{C}{3}-\frac{1}{2} (S+T)-\frac{i\sqrt{3}}{2}(S-T)}{D_m}}\\
\omega_3 & = &\displaystyle   \frac{1}{\Delta \sqrt{\epsilon \mu}}\sqrt{\frac{-\frac{C}{3}-\frac{1}{2} (S+T)+\frac{i\sqrt{3}}{2}(S-T)}{D_m}}
\ea
\ee
where $S$ et $T$ can be expressed in terms of $A,B$ and $C$:
\be
\ba{rcl}
S &=&\displaystyle (R + \sqrt{R^2 + Q^3})^{1/3}  \\
T &=&\displaystyle  (R - \sqrt{R^2 + Q^3})^{1/3}
\ea
\ee
with
\[
\ba{rcl}
R &=&\displaystyle \frac{9BC-27A-2C^3}{54 } \\[2ex]
Q &=&\displaystyle \frac{3B-C^2}{9}.
\ea
\]
The numerical phase velocities (noted $V'$) are obtained by dividing the roots $\omega_1, \omega_2$ and $\omega_3$ by $\beta$ (for the lossless regime, $\beta$ is equal to the wavenumber $k$, ; see equation \eqref{e_kdispersion} and \eqref{e_kgeneral})
\[
V'_1 \underset{\sigma=0}= \omega_1/k;  \; \; V'_2 \underset{\sigma=0}= \omega_2/k;\; \; V'_3\underset{\sigma=0}= \omega_3/k.
\]
The existence of three different relations for the phase velocities indicates that some numerical errors are generated by the introduction of finite-difference operators. We next define parameter $K$ as
\be
\label{e_Knorm}
K \equiv \frac{\beta \Delta}{2 \pi} \underset{\sigma=0}= \frac{k \Delta}{2 \pi} \,,
\ee
which represents the wavenumber in grid point unit. With this definition and the expression of the analytical phase velocity for the lossless regime (equation \eqref{e_AnalyticPhaseVelo} with $\sigma = 0$), we can write the normalized phase velocities as:
\be
\label{e_ST}
\ba{rcl}
\displaystyle \left[\frac{V'_1}{V}\right]_{\sigma=0} &=& \displaystyle \frac{1}{2\pi K}\sqrt{\frac{\frac{-C}{3} + S+T}{D_m}}; \\
\displaystyle  \left[\frac{V'_2}{V}\right]_{\sigma=0}  &=& \displaystyle  \frac{1}{2\pi K}\sqrt{\frac{\frac{-C}{3} -\frac{S+T}{2}-i\sqrt{3}(\frac{S-T}{2})}{D_m}}; \\
\displaystyle  \left[\frac{V'_3}{V}\right]_{\sigma=0} &=& \displaystyle  \frac{1}{2\pi K}\sqrt{\frac{\frac{-C}{3} -\frac{S+T}{2}+i\sqrt{3}(\frac{S-T}{2})}{D_m}}.
\ea
\ee
These three equations give the constraints to optimize the weighting coefficients $a$ and $b$ found in the finite-difference operators and the lumped operator. Since we need to find the coefficients for any propagation angle, we must express the finite-difference operators hidden in these last expressions as a function of the angles of propagation. With the initial harmonic substitution, the differential operators given in table \ref{t_DiffOperators} are
\be
\label{e_DiffAngles}
\ba{rcl}
 D_{xx}& =&\displaystyle -4\sin^2(\frac{k_x\Delta}{2}) \, \left [a + (1-a)\cos(k_z\Delta)\right ] \\
D_{zz}&= &\displaystyle  -4\sin^2(\frac{k_z\Delta}{2}) \, \left [a + (1-a)\cos(k_x\Delta)\right ] \\
D_{xz}&=&\displaystyle 1-{{\rm e}^{-i{\it k_x}\,\Delta}}-{{\rm e}^{-i{\it k_z}\,\Delta}}+{
{\rm e}^{-i\Delta\, \left( {\it k_x}+{\it k_z} \right) }} \\
D_{xz}^\star &=&\displaystyle  {{\rm e}^{i\Delta\, \left( {\it k_x}+{\it k_z} \right) }}-{{\rm e}^{i{
\it k_x}\,\Delta}}-{{\rm e}^{i{\it k_z}\,\Delta}}+1 \\
D_{x} &=& \displaystyle 1-{{\rm e}^{-i{\it k_x}\,\Delta}} \\
D_{x}^\star &=& \displaystyle {{\rm e}^{i{\it k_x}\,\Delta}}-1 \\
D_{z} &=&  \displaystyle {{\rm e}^{i{\it k_z}\,\Delta}}-1 \\
D_{z}^\star &=&\displaystyle  1-{{\rm e}^{-i{\it k_z}\,\Delta}}
\ea
\ee
and the lumped operator is
\be
\label{e_LumpAngles}
D_m =b + \frac{(1-b)}{2}\left (\cos \left( {\it k_z}\,\Delta \right) +\cos \left ({\it k_x}\,
\Delta \right) \right)
\ee
The components of the wavenumber $(k_x,k_z,k_y)$ can be expressed in terms of the propagation angles:
\[
\ba{rcl}
k_x &=&k \sin(\phi) \sin(\theta), \\
k_z &=&k \sin(\phi) \cos(\theta), \\
k_y &=&k \cos(\phi).
\ea
\]
With the definition of parameter $K$ (equation \eqref{e_Knorm}) and the fact that the wavenumber is real for the lossless regime ($k=\beta$), we can replace every occurrences of $k\Delta$ found in equations \eqref{e_ABC}, \eqref{e_DiffAngles} and  \eqref{e_LumpAngles} by
\be
\label{e_kdelta}
k \Delta =2 \pi K.
\ee
We can show that for $K \rightarrow 0$ (a high number of grid point per wavelength, a situation where the numerical errors vanishes), the normalize phase velocities simplify to
\[
\ba{rcl}
\displaystyle \left[\frac{V'_1}{V}\right]_{\substack{\sigma=0\\K \rightarrow 0}} &\rightarrow& 1, \\[3ex]
\displaystyle \left[\frac{V'_2}{V}\right]_{\substack{\sigma=0\\K \rightarrow 0}} &\rightarrow& 1, \\[3ex]
\displaystyle \left[\frac{V'_3}{V}\right]_{\substack{\sigma=0\\K \rightarrow 0}} &\rightarrow& 0.
\ea
\]
With these considerations, the normalized phase velocities $V'_1/V$ and $V'_2/V$ can be associated to some numerical anisotropy for the transverse electromagnetic wave introduced by the use of finite-difference operators. When $K$ increases, the normalized phase velocity $V'_3/V$ becomes different than zero, resulting in a propagation mode different than the purely transverse electromagnetic mode.  We can reduce the number of equation for the normalize velocities by imposing $V'_3/V = 0$ and by injecting this constraint into $V'_2/V'$. This operation reduces the number of constraint equations and we are left with two normalized phase velocities, one for each of the transverse oscillation mode. Using the notation $V^{\sigma=0}_{P1}$ for the normalized phase velocity $[V'_1/V]_{\sigma=0}$ and $V^{\sigma=0}_{P2}$ for $[V'_2/V]_{\sigma=0}$, these two normalized velocities are
\be
\label{e_V_Ph}
\ba{rcl}
V^{\sigma = 0}_{P1}&=& \displaystyle \frac{1}{2\pi K}\sqrt{\frac{\frac{-C}{3} + S+T}{D_m}}, \\
V^{\sigma = 0}_{P2}&=& \displaystyle \frac{1}{2\pi K}\sqrt{\frac{\frac{-2C}{3} -(S+T)}{D_m}}.
\ea
\ee
These two normalized phase velocities are the constraint equations.  Using  equations \eqref{e_ABC}, \eqref{e_ST}, \eqref{e_DiffAngles}, \eqref{e_LumpAngles} and \eqref{e_kdelta}, these two constraints equations will depend only on parameter $K$ (the wavenumber in grid points units), the propagation angles ($\theta$ and $\phi$) and the weighting coefficients $a$ and $b$. We can thus write
\[
\ba{rcl}
V^{\sigma = 0}_{P1} (K,\theta, \phi, a, b), \\[2ex]
V^{\sigma = 0}_{P2} (K,\theta, \phi, a, b).
\ea
\]
The normalized group velocities (noted $V^{\sigma = 0}_{G1}$ and $V^{\sigma = 0}_{G2}$) are given by
\be
\label{e_V_Gr}
\ba{rcl}
V^{\sigma = 0}_{G1}&=& \displaystyle \frac{1}{2\pi}\frac{\partial}{\partial K} \sqrt{\frac{\frac{-C}{3} + S+T}{D_m}}, \\
V^{\sigma = 0}_{G2}&=& \displaystyle \frac{1}{2\pi}\frac{\partial}{\partial K} \sqrt{\frac{\frac{-2C}{3} -(S+T)}{D_m}}.
\ea
\ee

\section{Heterogeneous formulation}

In this appendix, we give the finite-difference equation corresponding to equation \eqref{e_GeneralEtilde} for a heterogeneous medium and for the optimal weighting coefficients.  This equation is first written explicitly with additional coefficients $\xi_x$ and $\xi_z$ to include the absorbing boundary condition (see \cite{RAPPAPORT2000}):
\be
\label{e_HeteSource}
\left [ \ba{ccc}
c_{11} & c_{12}  & c_{13} \\
c_{21} & c_{22}  & c_{23} \\
c_{31} & c_{32}  & c_{33}
\ea \right ]
\left [\ba{c} \tilde E^x \\ \tilde E^z \\ \tilde E^y \ea \right ] =\left [\ba{c} J^x \\ J^z \\ J^y \ea \right]
\ee
with
\[
c_{11}= Y  - \frac{\partial_z}{\xi_z}\left(\frac{1}{\xi_zZ}\right )\partial_z  + \frac{k_y^2}{Z};
\]
\[
c_{22}= Y  - \frac{\partial_x}{\xi_x}\left(\frac{1}{\xi_xZ}\right )\partial_x  + \frac{k_y^2}{Z};
\]
\[
c_{33} = Y  - \frac{\partial_x}{\xi_x}\left(\frac{1}{\xi_xZ}\right )\partial_x  - \frac{\partial_z}{\xi_z}\left(\frac{1}{\xi_zZ}\right )\partial_z;
\]

\[
c_{12}=  \frac{\partial_z}{\xi_z}\left(\frac{1}{\xi_xZ}\right )\partial_x; \quad  c_{21} =  \frac{\partial_x}{\xi_x}\left(\frac{1}{\xi_zZ}\right )\partial_z;
\]
\[
c_{13} =\left(\frac{ik_y}{\xi_xZ}\right ){\partial_x}; \quad \quad \;\; c_{31} = \frac{ik_y}{\xi_x}\partial_x\left(\frac{1}{Z}\right );
\]
\[
c_{23} =  \left(\frac{ik_y}{\xi_zZ}\right ){\partial_z}; \quad \quad \;\;  c_{32} = \frac{ik_y}{\xi_z}\partial_z\left(\frac{1}{Z}\right ).
\]
Inside the absorbing region, the $\xi_\gamma$ coefficients are
\be
\xi_{\gamma} = 1 + A_m \left(1+i\frac{k_r}{k_i}\right)\left(\frac{d_\gamma}{T_\gamma}\right)^p, \; \; \; \gamma =\{x,z\}
\ee
where $k_r$ and $k_i$ are respectively the real and imaginary parts of the wavenumber when entering the absorbing region, $T_\gamma$ is the width of the absorbing region in the $\gamma$ direction and $d_\gamma$, the distance from the beginning of the absorbing region.  The coefficient $p$ is a positive real number between $3$ and $4$ \cite{RAPPAPORT2000} and $A_m$ is a parameter derived from a prescribed value of the reflection coefficient at the edge of the grid \cite{ELLEFSEN2009}. Outside the absorbing region, $\xi_{\gamma} = 1$.

We use the staggered grid given by Fig.~\ref{f_StaggeredGrid} and central differences with our extendedd $9$ points stencil to discretize the partial derivatives. The first line of equation \eqref{e_HeteSource} is discretized around the central point of the grid cell (around $J^x$). The second and third lines of this equation are discretized respectively around the top-right corner and right edge of the grid cell (around $J_z$ and $J_y$).

Using the notation $\tilde E^\gamma_{0,0}$ for the $\gamma = x, z,$ or $y$ components of $\bftE$ located on cell $i,j$ and  $\tilde E^\gamma_{\pm 1,\pm 1}$ for the same component on cell $i\pm1,j\pm1$, we can approximate partial derivatives in equation \eqref{e_HeteSource} with the extended difference operators. Since only the diagonal elements are modified by the optimal coefficients formulation, we give the explicit details only for these diagonal elements. The other terms are written with standard central differences and their development can be found in \cite{ELLEFSEN2009}.

The first diagonal element is calculated at the central point of the ${i,j}$ cell:
\be
\left[c_{11}\tilde E^x \right]_{i,j} \approx \mathcal{D}^x_{m} + \frac{k_y^2 E^x_{0,0}} {Z_{0,0}} - \frac{\mathcal{D}^x_{zz}}{\Delta_{z}^2}
\ee
with
\[
\ba{rcl}
\mathcal{D}^x_{m}&=&  bY_{0,0}E^x_{0,0} +  \frac{(1-b)}{4} \left(Y_{0,1}E^x_{0,1} + Y_{0,-1}E^x_{0,-1} \right) \nonumber \\[4pt]
 	 &	& \; \; \;+\frac{(1-b)}{4} \left(Y_{1,0}E^x_{1,0}+ Y_{-1,0}E^x_{-1,0} \right) \nonumber \\[4pt]
\mathcal{D}^x_{zz} &=& \displaystyle \sum_{s=-1}^{s=1}A^x_z(s)[E^x_{s,1}-E^x_{s,0}]-B^x_z(s)[E^x_{s,0}-E^x_{s,-1}]  \nonumber
\ea
\]
and where
\[
\ba{rcl}
A^x_z(s) &=& \displaystyle  \frac{a_{|s|}}{\xi_{z_{0,0}}(\xi_zZ)_{s,1/2}} \nonumber \\
B^x_z(s) &=& \displaystyle  \frac{a_{|s|}}{\xi_{z_{0,0}}(\xi_zZ)_{s,-1/2}}. \nonumber
\ea
\]
With this compact notation, $a_0=a$  ($a$ is the optimal coefficient of table 3) and $a_1 = (1-a)/2$. In coefficients $A^x_z(s)$ and $B^x_z(s)$, some parameters must be evaluated at intermediate (i.e. staggered) grid points and average expressions are then used. For instance, when calculating $A^x_z(s=0)$, the term $(\xi_zZ)_{0,1/2}$ is encountered and needs an evaluation at the right-edge of the cell. The following approximation is then used:

\be
\label{e_ApproxStag}
 \frac{1}{(\xi_zZ)_{0,1/2}} \approx \frac12\left ( \frac{1}{(\xi_zZ)_{0,0}} +\frac{1}{(\xi_zZ)_{0,1}} \right).
\ee

The second diagonal element is calculated at the top-right corner of the grid cell:
\be
\left[c_{22}E^z_{0,0}\right]_{i+1/2,j+1/2} \approx \mathcal{D}^z_{m} + \frac{k_y^2 E^z_{0,0}} {Z_{1/2,1/2}} - \frac{\mathcal{D}^z_{xx}}{\Delta_x^2}
\ee
with
\[
\ba{rcl}
\mathcal{D}^z_{m} &=&  bY_{1/2,1/2}E^z_{0,0} \nonumber \\[4pt]
& & \; \;  + \frac{(1-b)}{4} \left(Y_{1/2,1/2+1}E^z_{0,1} + Y_{1/2,1/2-1}E^z_{0,-1}\right ) \nonumber \\ [4pt]
&   &\;  \; \; \; + \frac{(1-b)}{4} \left(Y_{1/2+1,1/2}E^z_{1,0} + a_1Y_{1/2-1,1/2}E^z_{-1,0}\right ) \nonumber \\[4pt]
\mathcal{D}^z_{xx}  &=&\displaystyle  \sum_{t=-1}^{t=1}A^z_x(t)[E^z_{1,t}-E^z_{0,t}]-B^z_x(t)[E^z_{0,t}-E^z_{-1,t}] \nonumber
\ea
\]
where
\[
\ba{rcl}
A_{x}^z(t) &=& \displaystyle \frac{a_{|t|}}{\xi_{x_{1/2,1/2}}(\xi_xZ)_{1,1/2+t}} \nonumber \\[10pt]
B_{x}^z(t) &=&\displaystyle  \frac{a_{|t|}}{\xi_{x_{1/2,1/2}}(\xi_xZ)_{0,1/2+ t}}. \nonumber
\ea
\]
Once again, approximations similar to equation \eqref{e_ApproxStag} are used whenever parameters need to be evaluated at staggered grid points.

The last diagonal element is evaluated at the right edge of the grid cell:
\be
\left[c_{33}E^y_{0,0}\right]_{i+1/2,j} \approx \mathcal{D}^y_{m}- \frac{\mathcal{D}_{xx}^y}{\Delta^2_x} - \frac{\mathcal{D}_{zz}^y}{\Delta^2_z}
\ee
with
\[
\ba{rcl}
\mathcal{D}^y_{m}&=&  bY_{1/2,0}E^y_{0,0} +  \frac{(1-b)}{4} \left( Y_{1/2,1}E^y_{0,1} + Y_{1/2,-1}E^y_{0,-1} \right) \nonumber \\[4pt]
& & \;\;\;+ \frac{(1-b)}{4} \left(Y_{1/2+1,0}E^y_{1,0} + Y_{1/2-1,0}E^y_{-1,0}\right)  \nonumber \\[4pt]
\mathcal{D}_{xx}^y  &=& \displaystyle \sum_{t=-1}^{t=1}A_x^y(t)[E^y_{1,t}-E^y_{0,t}]-B_x^y(t)[E^y_{0,t}-E^y_{-1,t}] \nonumber \\ [4pt]
\mathcal{D}_{zz}^y  &=& \displaystyle \sum_{s=-1}^{s=1}A_z^y(s)[E^y_{s,1}-E^y_{s,0}]-B_z^y(s)[E^y_{s,0}-E^y_{s,-1}] \nonumber \\ [4pt]
\ea
\]
and where
\[
\ba{rcl}
A_{x}^y(t) &=& \displaystyle  \frac{a_{|t|}}{\xi_{x_{1/2,0}}(\xi_xZ)_{1,t}}  \\[10pt]
B_{x}^y(t) &=&\displaystyle  \frac{a_{|t|}}{\xi_{x_{1/2,0}}(\xi_xZ)_{0,t}} \\[10pt]
A_{z}^y(s) &=& \displaystyle \frac{a_{|s|}}{\xi_{z_{1/2,0}}(\xi_zZ)_{1/2+s,1/2}}  \\[10pt]
B_{z}^y(s) &=& \displaystyle  \frac{a_{|s|}}{\xi_{z_{1/2,0}}(\xi_zZ)_{1/2+s,-1/2}}. 
\ea
\]

\end{document}